\documentclass[9pt,twocolumn,twoside]{opticajnl}

\journal{optica} 

\setboolean{shortarticle}{false}

\usepackage{tikz}

\usepackage{lineno}

\title{Spectral phase interferometry for direct electric-field reconstruction of synchrotron radiation}

\author[1,*]{Takao Fuji} 
\author[2,3]{Tatsuo Kaneyasu} 
\author[3]{Masaki Fujimoto}
\author[3]{Yasuaki Okano} 
\author[3]{Elham Salehi} 
\author[4,5]{Masahito Hosaka} 
\author[4]{Yoshifumi Takashima} 
\author[4]{Atsushi Mano} 
\author[6]{Yasumasa Hikosaka} 
\author[7]{Shin-ichi Wada} 
\author[8,3,$\dagger$]{Masahiro Katoh} 

\affil[1]{Laser Science Laboratory, Toyota Technological Institute, 
Nagoya 468–8511, Japan
}
\affil[2]{SAGA Light Source, Tosu 841-0005, Japan}
\affil[3]{Institute for Molecular Science, Okazaki 444-8585, Japan
}
\affil[4]{Synchrotron Radiation Research Center, Nagoya University, Nagoya, 464-0814, Japan}
\affil[5]{National Synchrotron Radiation Laboratory, University of Science and Technology of China, Hefei 230029, China}
\affil[6]{Institute of Liberal Arts and Sciences, University of Toyama, Toyama 930-0194, Japan}
\affil[7]{Graduate School of Advanced Science and Engineering, Hiroshima University, Higashi-Hiroshima 739-8526, Japan}
\affil[8]{Hiroshima Synchrotron Radiation Center, Hiroshima University, Higashi-Hiroshima 739-0046, Japan}

\affil[*]{Corresponding author: fuji@toyota-ti.ac.jp}
\affil[$\dagger$]{Corresponding author: mkatoh@hiroshima-u.ac.jp}

\begin{abstract}
Ultraviolet and extreme ultraviolet electric-fields produced by relativistic electrons 
in an undulator of a synchrotron light source 
are characterized by using spectral phase interferometry for direct electric-field reconstruction (SPIDER). 
A tandem undulator with a phase shifter produces 
a pair of wavelength shifted wave packets with some delay. 
The interferogram between the pair of the wave packets    
is analyzed with a SPIDER algorithm, 
which is widely used for ultrashort pulse characterization. 
As a result, a 10-cycle square shaped electric-field is reconstructed. 
The waveform corresponds to the radiation from an electron accelerated 
with the undulator which consists of 10 periods of permanent magnets.
\end{abstract}

\setboolean{displaycopyright}{true}

\begin{document}

\maketitle

\section{Introduction}
Synchrotron light sources have been developed 
for more than 70 years 
and currently ultrashort X-ray pulses are generated with 
some free electron lasers (FELs)~\cite{NPhoto4-641,NPhoto6-540,NPhoto11-708,AS7-720,NPhoto14-391}. 
Characterization of the pulses generated from such sources  
is very challenging due to the short wavelength and 
the short pulse duration.

To characterize ultrashort pulses in the ultraviolet (UV) or extreme ultraviolet (XUV) region,  
it is straightforward to 
measure the cross-correlation 
between the test and reference pulses. 
However, it is very difficult to prepare a reference pulse 
for synchrotron light sources. 
Several synchronization systems between an ultrashort pulse laser 
and a synchrotron light source have been realized 
and used for the estimation of the duration of the pulses from some FELs~\cite{optica8-545,PTRS377-20180386,NPhoto12-215}, 
however, it is always challenging 
to synchronize such very different light sources within femtosecond timing jitter. 

\begin{figure*}[h]
\includegraphics[width=1.9\columnwidth]{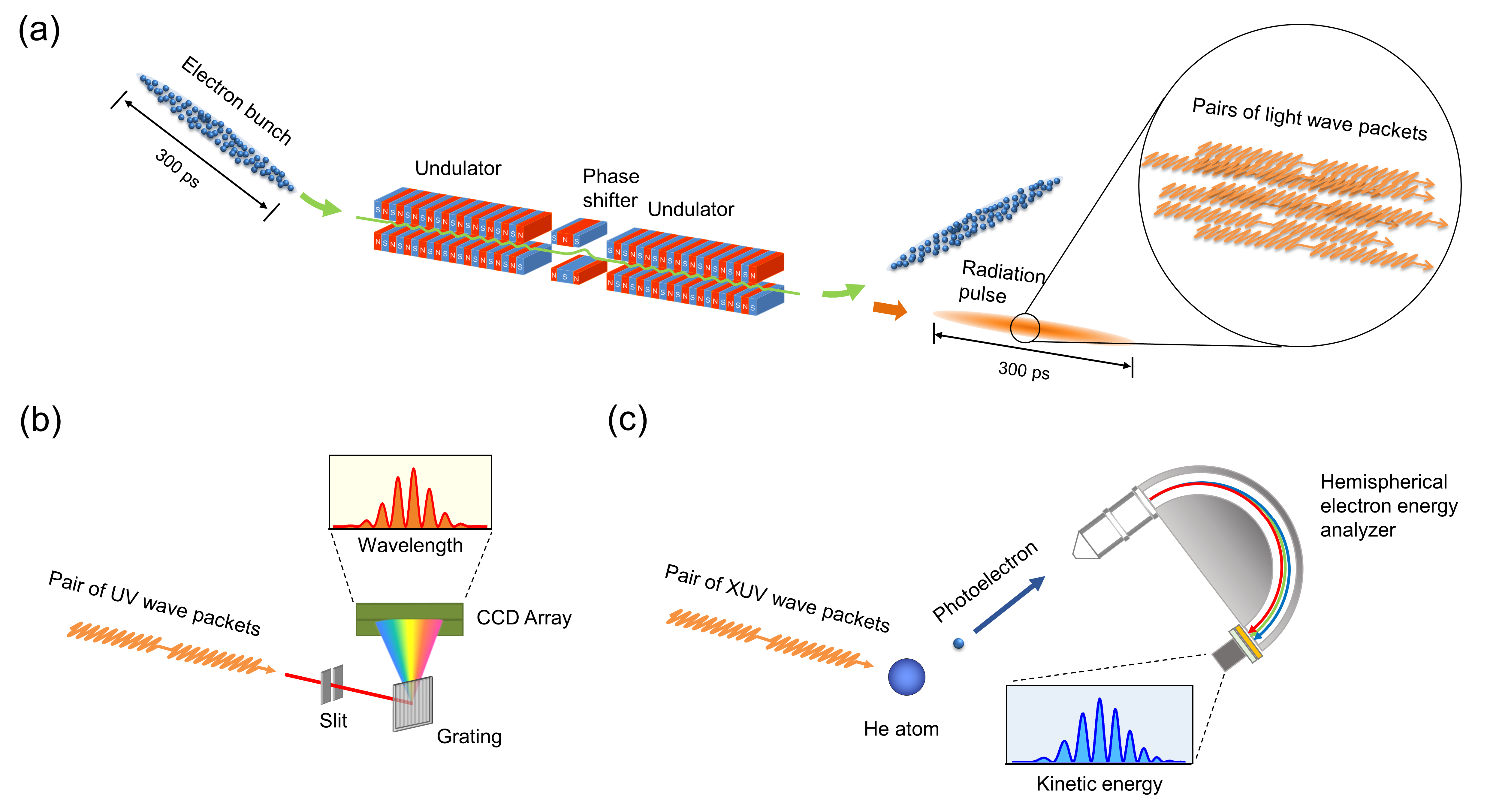}
\caption{\label{fig:setup}
Schematic of the experiment.
(a) Tandem undulator in the UVSOR-III synchrotron, consisting of two APPLE-II undulators and each relativistic electron in the bunch emits a pair of 10-cycle light wave packets. The undulators were set to horizontal linear polarization mode. The delay time between the wave packets is controlled by the phase shifter magnet between the two undulators. The light wave packets are randomly distributed within the overall pulse length of 300~ps (FWHM), reflecting the length of an electron bunch in the storage ring. (b) Setup for frequency-domain interferometry in the UV region. The UV spectrum of wave packets was measured by using a grating spectrometer. (c) Setup for photoelectron spectroscopy in the XUV region. The XUV spectra were derived from the photoelectron spectrum of helium. A hemispherical electron energy analyzer was used to measure the photoelectron spectrum. 
}
\end{figure*}

The waveform of the electric-field produced  
by an relativistic electron in the undulator 
is basically defined by 
the number of magnets and the gap between the magnets in the undulator. 
The number of permanent magnets and the gap between the magnets 
define the number of oscillations and the carrier wavelength of the waveform respectively. 
%
In the UVSOR-III synchrotron light source, there is a tandem undulator
which can produce two wave packets. The wavelength can be scanned from XUV
to visible region.
By changing the gap between the permanent magnets of each undulator, 
it is possible to change the wavelength of each wave packet individually. 
A phase shifter, 
which consists of three pairs of electromagnets and forms a small chicane for the electron beam, 
between the undulators 
can control the delay between the wave packets 
in femtosecond regime with an attosecond accuracy. 
The system was applied for coherent control of atoms and molecules~\cite{PRL123-233401,NCom10-4988,PRL126-113202}. 
It is important to characterize the electric-field 
produced in the undulator for such experiments. 


We reported linear interferometric autocorrelation measurements in the UV region for the spontaneous radiation from the tandem undulator of UVSOR-III recently~\cite{SREP12-9682}. 
The shapes of the measured autocorrelation traces were well reproduced by the calculations assuming that the wave packet had the form of a double-pulsed 10-cycle sinusoidal wave. 
However, the waveform of the wave packet cannot be directly reconstructed only by the autocorrelation trace. 
Thus, for accurate measurement of the waveform of the electric field emitted by a single electron passing through the undulator, it is essential to introduce a pulse characterization method which allows for determining both the spectral phase and amplitude of the light pulse.

In this paper, we report the characterization of the electric-field waveform 
generated in the undulator using an algorithm of spectral phase interferometry 
for direct electric-field reconstruction (SPIDER)~\cite{OL23-792,JQE35-501}. 
To our knowledge, it is the first time to characterize 
the UV and XUV electric-field produced by an accelerated electron 
in the undulator 
without assuming the shape of the waveform. 

\section{SPIDER of synchrotron radiation}

SPIDER is a well-established femtosecond pulse characterization method 
invented in 1998~\cite{OL23-792}. 
The concept of the SPIDER is  
the retrieval of the spectral phase 
by analyzing the fringes 
of the interferogram between the test pulse 
and a spectrally sheared replica. 
In order to obtain such a pair of pulses,  
a strongly chirped pulse is prepared 
and sum frequency between the chirped pulse and the test pulse 
at two different delay times is taken. 
In this way, two delayed pulses 
with slightly different center wavelengths, 
namely spectrally sheared replicas of the original pulse, 
are obtained. 
The fringe deviation of the interferogram 
from that of two delayed pulses with the same wavelength, namely zero sheared replicas,   
corresponds to the derivative of the spectral phase with the frequency, 
namely, the group delay. 
By integrating the group delay with the frequency, 
the spectral phase is obtained 
and by calculating the Fourier transform of it together with the power spectrum 
it is possible to reconstruct the time-domain picture of the electric-field 
of the test pulse. 


The twin tandem undulator generates a pair of wavelength shifted wave packets with some delay. 
The interferogram between the pair of the wave packets can be considered as a ``SPIDER'' interferogram. 
Therefore, we can apply the same algorithm as the SPIDER 
to the interferogram 
to reconstruct the electric-field generated 
from the undulator. 


The important difference from the electric-field reconstruction of ultrashort laser pulses 
is that the synchrotron radiation is basically incoherent. 
In general, many electrons ($\sim10^9$ in the current case) compose an electron bunch with the duration of a few hundred picoseconds,  
and are circulating together in the storage ring. 
Each electron produces a wave packet of light which does not interfere with each other. 
As a result, the generated light pulse consists of randomly superimposed $\sim10^9$ wave packets 
within the duration of a few hundred picoseconds. 
However, each wave packet is supposed to be identical with each other.  
Thus, by using the SPIDER analysis, we obtain the electric-field of the wave packet generated from a single electron. 
For example, at BL1U of UVSOR-III a 10-cycle wave packet, 
corresponding to $\sim$1.2~fs pulse when it is operated at 35~eV, 
is produced. 
We aim to characterize such a waveform.  


\section{Experimental}

The experiment was carried out at the 750-MeV UVSOR-III storage ring ~\cite{JPCS425-042013}. Figure~\ref{fig:setup}(a) shows a schematic illustration of the tandem undulator installed in the UVSOR-III storage ring. The tandem undulator consisted of twin APPLE-II type variable polarization devices which were operated in the horizontal linear polarization mode. The number of magnetic periods, and the period length of the undulators are 10 and 88 mm respectively. 
We performed experiment at two different wavelength ranges, 
UV and XUV. 
The central photon energy of the fundamental radiation was adjusted to 3.5~eV and 35~eV, respectively. 

The spectral width of the radiation pulse was about 10\% of the central frequency, which translates to about 10 periods in the time domain. The electron beam current was 30~mA, which corresponds to $10^9$ electrons per bunch. Each relativistic electron in the bunch that passes through the tandem undulator emits a pair of wave packets 
whose waveforms are expected as time-separated 10-cycle oscillations, with longitudinal coherence between them. 
The spacing between the wave packets was tuned with attosecond precision by the phase shifter magnet that controls the electron path length between the undulators, 
this technique is used for a beam energy measurement~\cite{PRSTAB13-080702} or the cross undulator system~\cite{JSR21-352}.

Figure~\ref{fig:setup}(b) shows the experimental setup used for the UV wavelength range.
When we characterized the UV light wave, the undulator radiation was extracted into the air through an Al$_2$O$_3$ optical window. 
The spectrum of the wave packet was recorded with a grating spectrometer 
for the wavelength range from 200 to 400~nm with the resolution of 0.27~nm. 
The central part of the undulator radiation was introduced into the
entrance slit of the spectrometer 
located 7~m downstream from the middle point of two undulators.


For characterizing the XUV light wave, we measured the photoelectron spectrum of helium by the irradiation of the light (see Fig.~\ref{fig:setup}(c)). The photoelectron spectrum essentially reflects the spectral profile of the ionizing light, as the photoionization is proportional to the power spectral density of the double-pulsed ionizing light field~\cite{PRL89-173001, Optcom264-285}. Compared to an optical method using a grating spectrometer, photoelectron spectroscopy has the advantage of applicability to XUV or shorter wavelength light.  
In the measurement, the central part of the undulator radiation was cut out by a 0.4-mm-diameter pinhole located 9~m downstream from the middle point of two undulators. 
After passing through the pinhole, a toroidal mirror focused the undulator radiation into the gas cell of a hemispherical electron energy analyzer. 
The observation angle of photoelectrons was set to 55 degrees with respect to the polarization vector of the XUV light. 
The energy resolution of the analyzer was approximately 30~meV. 
The spectra of the XUV light is derived 
by correcting the measured photoelectron spectra 
with the photon energy dependence of the photoionization efficiency obtained from the cross section data~\cite{JES71-205}.

It is worth mentioning the time-domain view of the photoionization of an atom on the interaction with a pair of XUV wave packets. A pair of XUV wave packets produces two photoelectron wave packets in free space. The photoelectron wave packets in pairs spread in free space due to the dispersion and overlap each other, leading to the appearance of interference fringes in the spatial distribution. This wave packet interference can be observed as a fringe pattern in the energy or momentum distributions of photoelectrons~\cite{PRL89-173001, Optcom264-285}. The fringed profiles in the measured photoelectrons spectra are thus obtained.

\section{Electric-field reconstruction}

\begin{figure}
\includegraphics[width=\columnwidth]{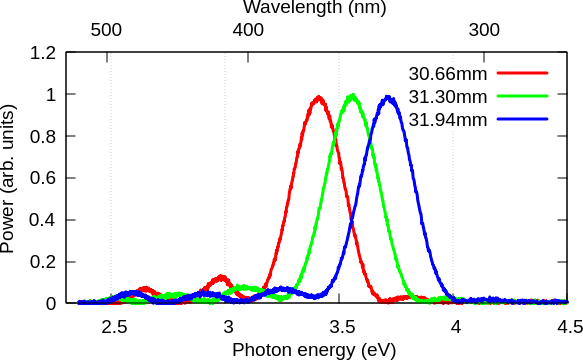}
\caption{\label{fig:spectra} UV power spectra of the synchrotron radiation from the undulator in the downstream.
Not to generate any light from the up stream, the gap of the upper undulator was open (200.0~mm).}
\end{figure}

The UV spectra of the synchrotron radiation from the undulator in the downstream are shown in Fig.~\ref{fig:spectra}. 
The wavelength of the UV wave packet can be tuned by changing the gap of the permanent magnet. 
The squared sinc-function ($\sin^2{x}/x^2$) like spectra indicates that the envelope of the wave packet is rectangle.  
In this experiment, the gap was changed from 30.66 to 31.94~mm, 
which results that the center wavelength of the wave packet was tuned from 365 to 334~nm respectively.
We fixed the gap of the upstream undulator to 31.30~mm corresponding to the center wavelength of 350~nm 
and scanned the gap of the downstream undulator for having a spectrally sheared replica. 

In the case of the XUV experiment, 
we fixed the gap of the upstream undulator to 74~mm corresponding to the center photon energy of 35~eV 
and scanned the gap of the downstream undulator for having a spectrally sheared replica.

\subsection{Electric-field reconstruction of a rectangular envelope pulse}

Here we would like to discuss the importance of the phase retrieval to characterize a rectangular envelope pulse. 
As can be seen in Fig.~\ref{fig:spectra}, 
the measured spectrum is similar to a squared sinc-function 
and it is easy to predict a rectangular envelope waveform in time-domain. 
However, simple Fourier-transformation of the spectrum (assuming a flat phase) never reproduces the right shape in time-domain. 
The phases of the oscillating components in the tail of the spectrum flip by $\pi$ next to each other. 
Those phase flips are essential to reproduce the rectangular envelope of the pulse. 
Therefore, the spectral phase measurement is very important to discuss the time-domain waveform of the pulse 
which has such a spectrum. 

In fact, the estimation of the phase difference between spectrally separated components 
is a challenge for ultrashort pulse characterization because the phase information is lost at the zero intensity region. 
Even by using SPIDER it is not straightforward.  
It is necessary to use a so large shear that the interference between the two separated components can be recorded, 
however, the resolution of the retrieved spectral phase becomes worse with a larger shear. 
To avoid the trade off, multi-shear algorithm is very useful~\cite{JOSAB26-1818}. 
The concept of the multi-shear algorithm is that SPIDER spectra with several different shears are recorded 
and the spectral phase of the target pulse is obtained to be consistent for all the shears in the least-square sense. 
Here we have adopted the method to reconstruct the electric-field of the synchrotron radiation.

\subsection{Experimental results}

\begin{figure}
\begin{tikzpicture}
\draw (0,7) node [above right]{\includegraphics[scale=0.51]{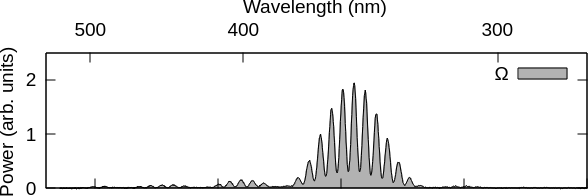}};
\draw (0,4.9) node [above right]{\includegraphics[scale=0.51]{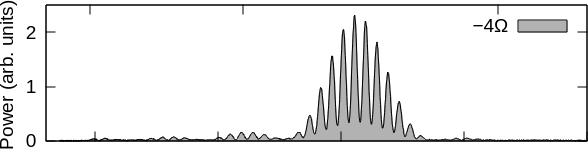}}; 
\draw (0,2.8) node [above right]{\includegraphics[scale=0.51]{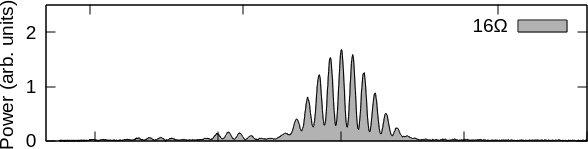}};
\draw (0,0) node [above right]{\includegraphics[scale=0.51]{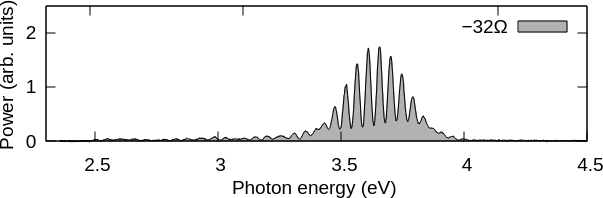}};
\draw (1.1,8.8) node [font=\sffamily] {(a)};
\draw (1.1,6.7) node [font=\sffamily] {(b)};
\draw (1.1,4.6) node [font=\sffamily] {(c)};
\draw (1.1,2.4) node [font=\sffamily] {(d)};
\end{tikzpicture}
\caption{\label{fig:sheared} UV power spectra of the pairs of the wave packets generated from the twin undulator. 
The delay between the two wave packets was set to 90.25~fs. 
The frequency shift of the downstream wave packet is
(a) $\Omega$, (b) $-4\Omega$, (c) $16\Omega$, and (b) $-32\Omega$, 
where $\Omega = 2 \pi \times 1.15$~THz.}
\end{figure}
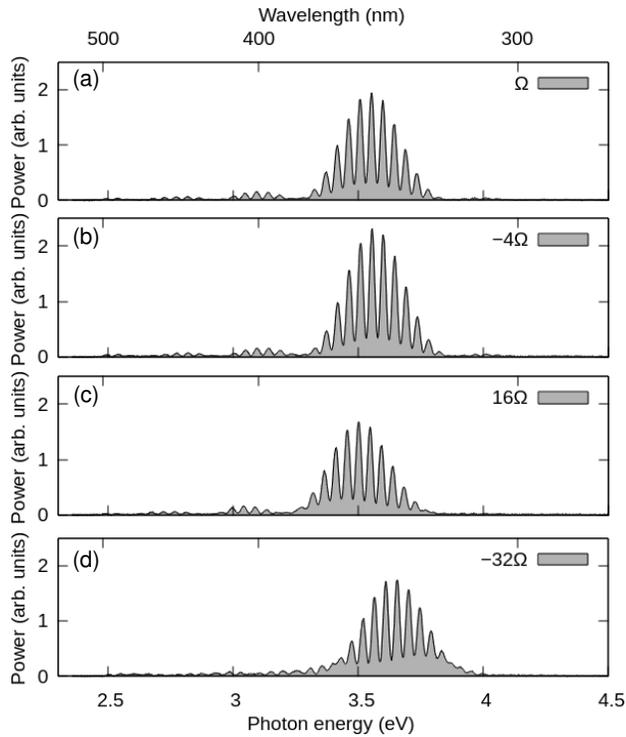
\begin{figure}
\begin{tikzpicture}
\draw (0,5) node [above right] {%
\includegraphics[width=\columnwidth]{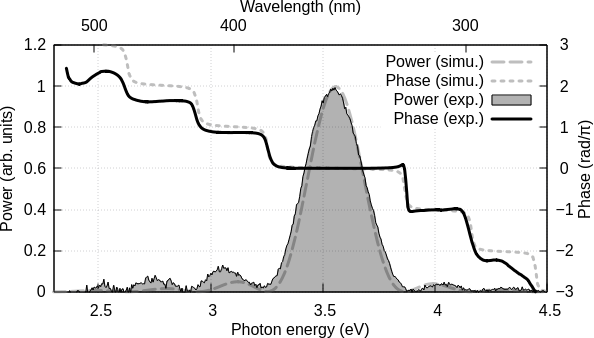}};
\draw (0,0) node [above right] {%
\includegraphics[width=\columnwidth]{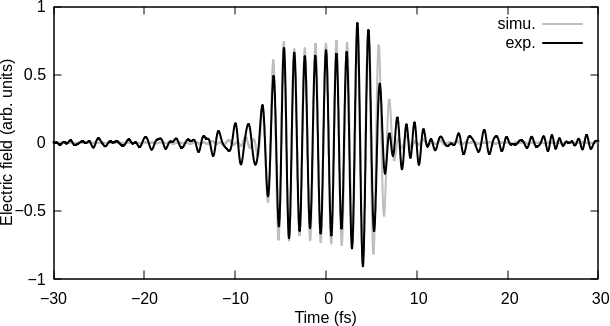}};
\draw (1.5,8.5) node [font=\sffamily] {(a)};
\draw (1.5,4.2) node [font=\sffamily] {(b)};
\end{tikzpicture}
\caption{\label{fig:retrieved} (a) UV power spectrum of the synchrotron light generated from the upstream undulator 
where the gap is 31.30~mm (shaded curve) and the spectral phase retrieved from the multi-shear algorithm. The numerically simulated power spectrum and spectral phase are also shown as a gray dashed curve and a gray dotted curve, respectively.  
The phase offset is set to zero at 3.6~eV. 
(b) Waveform of the wave packet retrieved from the power spectrum and spectral phase shown in (a). The numerically simulated waveform is also shown as a gray curve.}
\end{figure}

The UV spectra of the synchrotron light generated from the tandem undulator 
with several different gaps of the downstream undulator are shown in Fig.~\ref{fig:sheared}. 
The delay between the two wave packets is controlled by using the phase shifter between the two undulators. 
By recording a reference interferogram with zero shear, 
the delay between the two wave packets is estimated as 90.25~fs. 
The minimum shear angular frequency $\Omega$ is $2\pi \times 1.15$~THz (4.76~meV), 
which defines the frequency resolution of the measurement. 
The recorded shear frequencies are $\pm 2^n \Omega$ where $n$ is from 0 to 5. 
We measured 50 spectra for each shear and analyzed them with the multi-shear algorithm. 
The averaged spectral phase and the power spectrum of the upstream undulator beam 
is shown in Fig.~\ref{fig:retrieved}(a). 
The $\pi$ phase flips at each (nearly) zero-crossing points of the spectra 
are clearly seen, which corresponds to the step function like spectral phase. 
The phase offset is set to zero at the carrier frequency, 3.6~eV, 
since it cannot be determined with this scheme.  
The retrieved electric-field is shown in Fig.~\ref{fig:retrieved}(b). 
The rectangular envelope of the wave packet is very well-retrieved. 

\begin{figure}
\begin{tikzpicture}
\draw (0,5) node [above right] {%
\includegraphics[width=\columnwidth]{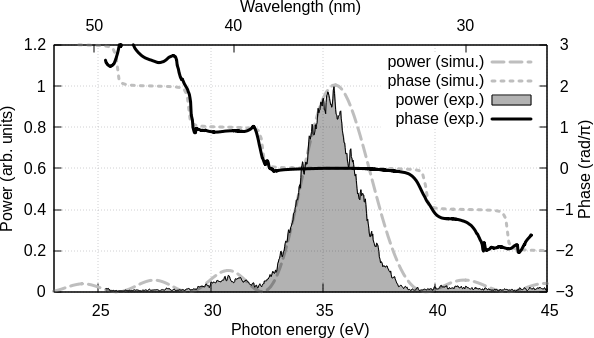}};
\draw (0,0) node [above right] {%
\includegraphics[width=\columnwidth]{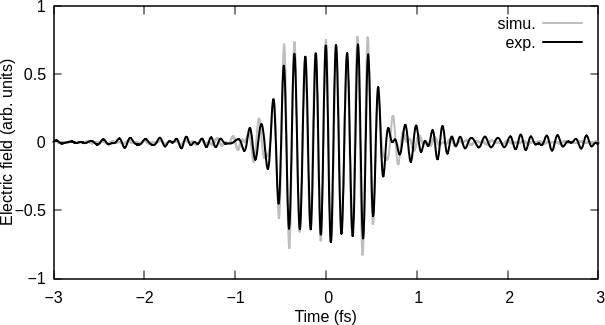}};
\draw (1.5,8.5) node [font=\sffamily] {(a)};
\draw (1.5,4.2) node [font=\sffamily] {(b)};
\end{tikzpicture}
\caption{\label{fig:retrievedXUV} (a) XUV power spectrum of the synchrotron light generated from upstream undulator 
where the gap is 74~mm (shaded curve) and the spectral phase retrieved from the multi-shear algorithm. The numerically simulated power spectrum and spectral phase are also shown as a gray dashed curve and a gray dotted curve, respectively. 
The phase offset is set to zero at 35~eV. 
(b) Waveform of the wave packet retrieved from the power spectrum and spectral phase shown in (a). The numerically simulated waveform is also shown as a gray curve.}
\end{figure}

We performed basically the same experiment for XUV. 
The delay between the two wave packets is estimated as 9.00~fs. 
The minimum shear angular frequency $\Omega_X$ corresponds to a photon energy of 27~meV. 
The recorded shear angular frequencies are $\pm\Omega_X$, $\pm2\Omega_X$, $\pm5\Omega_X$, $\pm11\Omega_X$, and $\pm23\Omega_X$. 
We measured a photoelectron spectrum for each shear 
and analyze them with the multi-shear algorithm. 
The spectral phase and the power spectrum of the upstream undulator beam 
is shown in Fig.~\ref{fig:retrievedXUV}(a). 
The $\pi$ phase flips at each (nearly) zero-crossing points of the spectra 
are also clearly seen. 
The phase offset is set to zero at the center photon energy, 36~eV.  
The retrieved electric-field is shown in Fig.~\ref{fig:retrievedXUV}(b). 
The rectangular envelope of the wave packet is very well-retrieved. 

We performed numerical simulations for the electric-field generation from a relativistic electron. 
We calculated the electron trajectories in the undulator magnetic field estimated from the actual arrangement of the permanent magnets.  
The time domain picture of the electric-field emitted by the electron was computed by means of Lienard-Wiechert field~\cite{jackson_classical_1999}. 
Performing Fourier transform on the electric field, 
we obtained power spectra and spectral phases shown as gray curves in Fig.~\ref{fig:retrieved}(a) and Fig.~\ref{fig:retrievedXUV}(a). 
Filtering out the higher harmonic components 
and performing inverse Fourier transform, 
we obtained the waveforms in time domain shown as gray curves in Fig.~\ref{fig:retrieved}(b) and Fig.~\ref{fig:retrievedXUV}(b). 
The $\pi$ phase flips at the zero-crossing points of the power spectra 
and the 10-cycle rectangular envelope waveforms in time-domain are consistent with the experimental results. 

Here we would like to discuss where the measured point of the electric-field is. 
In the case of laser pulse characterization with SPIDER, 
the recorded waveform reflects the waveform where the sheared replica is produced, 
namely, where the nonlinear wavelength conversion takes place~\cite{OL29-210}. 
In the current condition, the sheared replica is produced 
at the downstream undulator, 
therefore, we are supposed to measure the upstream wave packet 
at the time when the other wave packet generated in the downstream undulator. 
This is the reason why we do not observe any chirp on the recorded electric-field 
even though the UV light passes through several meters of air and several millimeter thick windows.

\section{Conclusion}

In conclusion, we have succeeded in the measurement of the electric-field of the synchrotron radiation 
by using multiple-shearing SPIDER algorithm. 
10-cycle rectangular envelope UV and XUV electric-fields are retrieved. 
The pulse durations of the UV and XUV wave packets were estimated as 12~fs and 1.2~fs, respectively.
The shape is predicted by a simple theory based on the interaction between a magnetic field and a relativistic electron. 
The result indicates that each electron in the bunch produces exactly the same electric field. 
This feature is one of the largest difference from standard incoherent light such as radiation from an incandescent lamp. 
If each electron in the bunch produces a different wave packet with each other, 
the SPIDER interferogram never shows up. 
This is a strong experimental evidence of the coherence of the synchrotron radiation.  
Since this method does not need nonlinear wavelength conversion, 
it would be very useful for shorter wavelength synchrotron radiation such as soft and hard X-ray. 
In particular, the method can immediately be used  
for two-color free electron lasers~\cite{NCom4-2919}. 

The pulse duration of a hard X-ray pulse from SACLA~\cite{NPhoto6-540} has just been estimated using an autocorrelation measurement based on two-photon absorption with a tandem undulator~\cite{PRR4-L012035}. 
It takes $\sim$37 hours to record a single autocorrelation trace 
because of the weak nonlinear signal in this wavelength range   
and it is necessary to assume the shape of the pulse 
for the estimation of the pulse duration. 
We expect that the SPIDER method can dramatically change the situation 
since any nonlinear effects are not used. 
The signal can be so large that single shot measurements are possible~\cite{JSR29-862,PRL109-144801}. 
In addition, we are able to characterize the waveforms of the XFEL pulses 
without any assumption of the pulse shape.  
Such a dramatic improvement of the XFEL pulse characterization 
has a great impact not only on the development of the XFEL 
but also on their applications, such as attosecond science, high-field physics, protein crystallography.

\begin{backmatter}
\bmsection{Funding} 
This work was supported in part by JSPS KAKENHI (grant numbers 20H00164, 22H02044).

 \bmsection{Acknowledgments} 
 The experiments were performed at the BL1U of UVSOR Synchrotron 
Facility, Institute for Molecular Science (21-805 and 22IMS6611). 
The construction of BL1U at UVSOR was supported by the Quantum Beam Technology Program of MEXT/JST. 

\bmsection{Disclosures} The authors declare no conflicts of interest.








\bmsection{Data availability} Data underlying the results presented in this paper are not publicly available at this time but may be obtained from the authors upon reasonable request.



\end{backmatter}





\bibliography{references}

\bibliographyfullrefs{references}


\ifthenelse{\equal{\journalref}{aop}}{%
\section*{Author Biographies}
\begingroup
\setlength\intextsep{0pt}
\begin{minipage}[t][6.3cm][t]{1.0\textwidth} 
  \begin{wrapfigure}{L}{0.25\textwidth}
    \includegraphics[width=0.25\textwidth]{john_smith.eps}
  \end{wrapfigure}
  \noindent
  {\bfseries John Smith} received his BSc (Mathematics) in 2000 from The University of Maryland. His research interests include lasers and optics.
\end{minipage}
\begin{minipage}{1.0\textwidth}
  \begin{wrapfigure}{L}{0.25\textwidth}
    \includegraphics[width=0.25\textwidth]{alice_smith.eps}
  \end{wrapfigure}
  \noindent
  {\bfseries Alice Smith} also received her BSc (Mathematics) in 2000 from The University of Maryland. Her research interests also include lasers and optics.
\end{minipage}
\endgroup
}{}

\end{document}